\documentclass[5p]{elsarticle}

\usepackage{lineno,hyperref}
\modulolinenumbers[5]
\usepackage{graphicx}
\usepackage{dcolumn}
\usepackage{bm}
\usepackage{float}
\usepackage{amsmath, amssymb}
\usepackage{xcolor}
\usepackage{widetext}

\journal{Chaos, Solitons \& Fractals}

\begin{document}

\begin{frontmatter}

\title{Order symmetry breaking and broad distribution of events in spiking neural networks with continuous membrane potential}


\author[1]{Marco Stucchi}

\author[2]{Fabrizio Pittorino\corref{mycorrespondingauthor}}
\cortext[mycorrespondingauthor]{Corresponding author}
\ead{fabrizio.pittorino@unibocconi.it}

\author[3]{Matteo di Volo}
\author[4,5]{Alessandro Vezzani}
\author[5,6]{Raffaella Burioni}

\address[1]{Department of Biology, University of Konstanz, Universit\"atsstrasse 10, 78457, Konstanz, Germany}

\address[2]{Artificial Intelligence Lab, Institute for Data Science and Analytics, Bocconi University, Milano, Italy}

\address[3]{Laboratoire de Physique Th\'eorique et Modelisation, Universit\'e de Cergy-Pontoise, 95302 Cergy-Pontoise cedex, France}

\address[4]{IMEM-CNR, Parco Area delle Scienze, 37/A-43124 Parma, Italy}

\address[5]{Dipartimento di Scienze Matematiche, Fisiche e Informatiche, Universit\`a  di Parma,
Parco Area delle Scienze 7/A  - 43124, Parma, Italy}

\address[6]{INFN, Gruppo Collegato di Parma, Parco Area delle Scienze 7/A - 43124, Parma, Italy}

\begin{abstract}
We introduce an exactly integrable version of the well-known leaky integrate-and-fire (LIF) model, with  continuous membrane potential at the spiking event, the \emph{c-LIF}. We investigate the dynamical regimes of a fully connected network of excitatory c-LIF neurons in the presence of short-term synaptic plasticity. By varying the coupling strength among neurons, we show that a complex chaotic dynamics arises, characterized by scale free avalanches.  The origin of this phenomenon in the c-LIF can be related to the order symmetry breaking in neurons spike-times, which corresponds to the onset of a broad activity distribution. Our analysis uncovers a general mechanism through which networks of simple neurons can be attracted to a complex basin in the phase space.
\end{abstract}

\begin{keyword}
neuron dynamics \sep synchronization \sep oscillations \sep criticality \sep chaos \sep symmetry breaking
\MSC[2010] 00-01\sep  99-00
\end{keyword}

\end{frontmatter}


\section{Introduction}
Networks of living neurons exhibit complex collective patterns of activity, ranging from asynchronous to synchronous firing even in absence of stimuli \cite{vree96,fox07}. The different degree of synchrony associated with these regimes is a fundamental property of the network dynamics and it is at the basis of the collective oscillatory rhythms observed in neural activity in the cortex \cite{buzsaki2012mechanisms}. A particularly interesting dynamical regime on the border between synchronous and asynchronous  dynamics is represented by the outbursts of activity called  \emph{neural avalanches} \cite{Beggs03122003,peterman09,bellay2015irregular}. These are cascades of activity clustered in time and separated by periods of quiescence and they have been observed both \emph{in vitro}  and \emph{in vivo} in different cortical areas and across many species \cite{Beggs03122003,persi2004modeling,bellay2015irregular,shriki2013neuronal}. Neuronal avalanches represent a peculiar mode of activity in the cortex, whose function and origin is still much debated \cite{buzsaki2006rhythms}. Their name stems from the fact that the distribution of the number of spiking neurons, i.e the size of the outburst, and the distribution of their durations lack a characteristic scale and are compatible with power laws, whose exponents fulfill scaling relations \cite{Friedman12,munoz99}. The dynamical activity in the avalanches regime also exhibits peculiar correlation properties and optimal computational performance \cite{mora2011biological,munoz18,wilting18}. These findings suggest that neural networks might operate close to a critical point, whose type and origin is still debated  \cite{Chialvo2010} and could be associated to a hybrid synchronization regime \cite{disanto18,buendia20}.

Another open point is what are the minimal microscopic ingredients at the level of neuron dynamics that can account for such an emergent dynamical regime. In a recent paper \cite{Pittorino} we have shown that an avalanches dynamical regime, with peculiar synchronization and correlation properties, can emerge in a sparse heterogeneous (in the sense of neuronal connectivity) network of leaky-integrate and fire (LIF) neurons \cite{Brunel2007trad, Brunel2007} with short-term synaptic plasticity modeled through the Tsodyks-Uziel-Markram mechanism \cite{TsodyksMarkram, TsodyksUzielMarkram}.  Interestingly, this regime emerges from a transition to chaos observed in the corresponding homogeneous network (all-to-all connections), which turns into an avalanches regime in the presence of heterogeneous connectivities, suggesting  a fundamental role of neuronal heterogeneity. While the primary role of excitatory connections and a regulatory mechanism, such as synaptic plasticity, have been shown to be key ingredients, it is not understood to what extent this scenario depends on the structural organization of neuron connections and on the  choice of single neuron intrinsic dynamics. 

In particular, even if the LIF model has the great advantage of being analytically integrable, it suffers from a discontinuity at the spiking time that may have strong effects on a dynamical regime like the one we are studying, where  neurons have a high synchronous activity during the burst. In order to investigate the relevance of single neuron time scales, we introduce a new LIF model which is analytically integrable and  removes the  hard discontinuity in voltage dynamics at the spiking event, thus mimicking the continuous dynamic of the action potential (see \cite{IzhikevichB} for a discussion of models with continuous membrane potential). This requires an additional variable which, using the physics jargon, plays the role of a mass in the context of a second order differential equation. According to its continuous features, we call such model the  \emph{c-LIF} (continuous LIF).

We study a homogeneous fully connected network of excitatory c-LIF neurons with short-term synaptic plasticity and we show that it exhibits a complex bursting dynamical regime with characteristic scale free dynamics, even without heterogeneity in the network. The microscopic inertial timescale of the c-LIF has a desynchronizing effect leading to avalanches of activity when acting in the region of parameters where the homogeneous network has a chaotic regime. By using a mean field approach, in this simple model we can trace back the mechanism for the onset of the broad activity distributions and show that the avalanche dynamical regime results from an order symmetry breaking, i.e. from neurons crossings in the sequence of firing times of the network. The possibility to break this symmetry has already been investigated in \cite{kirst2009sequential, timmesiam} in the presence of partial resets. Here we show that order symmetry breaking is a crucial ingredient to produce bursts with an internal structure and scale free avalanches distributions in a homogeneous network, with peculiar syncronization properties. This also clarifies that networks with homogeneous connections can give rise to an avalanche regime, even if neurons are all identical, both in their intrinsic dynamics and in the way they are connected to the surrounding network.

The article is organized as follows. In Sec.~\ref{sec:model} we briefly recall the LIF equation and we extend it to the c-LIF single neuron model. We then
consider $N$ globally coupled excitatory c-LIF neurons, with the dynamics of their interactions following the TUM model for short-term synaptic plasticity, accounting for the synaptic transmitter dynamics. 
In Sec. \ref{sec:taum_0} we recall the dynamical features of networks of LIF neurons with short-term synaptic plasticity and in 
Sec. \ref{sec:dynamics} we investigate the dynamics of the c-LIF fully connected network in the parameter space, observing quasi-synchronous, asynchronous and bursty chaotic dynamics. 
In Sec. \ref{sec:symmetry} we link the complex collective dynamics to a symmetry breaking of the spiking order of neurons. 
In Sec. \ref{sec:synchronization} we study the synchronization properties of the model.
Finally, Sec. \ref{sec:conclusions} is reserved for discussions and future perspectives.

\section{\label{sec:model} The c-LIF model}

The LIF neuron model \cite{dayanabbott} retains the basic features of a biological neuron, while being  easy to analyse mathematically and computationally. The dynamical equation  is:
\begin{equation}
	\tau_{1}\dot{V}_{i}=E_{c}-V_{i}+R_{in}GI_{i}, \label{eq:lif}
\end{equation}
where $\tau_{1}$ is the membrane time constant, $R_{in}$ is the
membrane resistance, $GI_{i}$ is the synaptic current received
by neuron $i$ from all its pre-synaptic neurons ($G$ is a parameter regulating its strength) and $E_{c}$ is the contribution of a constant external current (multiplied by $R_{in}$).
Whenever the potential $V_{i}(t)$ reaches the threshold value $V_{th}$
it is reset to $V_{r}$ and a spike is sent to all post-synaptic neurons.
It is useful to rescale time with the membrane time constant $\tau_{1}$
and to introduce the dimensionless quantities $a=\frac{E_{c}-V_{r}}{V_{th}-V_{r}}$, $g=\frac{R_{in}G}{V_{th}-V_{r}}$, $v=\frac{V-V_{r}}{V_{th}-V_{r}}$,  so that $v_{r}=0$ and $v_{th}=1$.  
With this rescaling, the difference between the potential at the threshold and at the reset is set to $1$ and the unit of $v$ is of the order of $10 mV$ (considering $V_{th}=-55 mV$ and $V_{r}=-65 mV$). The membrane potential $v_{i}$ of a neuron $i$ with an interaction current  $I_i(t)$ evolves in time according to the differential equation: 
\begin{eqnarray}
	\tau_{1}\dot{v}_{i}(t) & = & a - v_{i}(t) + g I_i \left(t\right). \label{eq:vk}
\end{eqnarray}
The potential is reset to $0$ at times $t_{i}(m)$ when it reaches the threshold $v_{i}(t_{i}(m))=1$.  

The LIF model describes with an instantaneous reset of the voltage $V$ a process that in reality determines the continuous and non-instantaneous rise and the decay of the membrane voltage $V$ during the emission of a spike.  This continuous dynamics of the voltage $V$  at the spike emission has been explained by the seminal work of Hodgkin and Huxley \cite{hodgkin1952quantitative}, through a four dimensional set of differential equations accounting for the kinetics of  potassium and sodium channels in neurons' membranes. In the direction of simplifying this description, but still accounting for its fundamental features, many models have been proposed. Some of them were able to reduce the dimensionality of the system  and to keep track of the voltage dynamics at the spike emission, as in the Morris Lecar \cite{morris1981voltage} or  FitzHugh-Nagumo models \cite{fitzhugh1961impulses}. These models are nevertheless not integrable, at variance with the LIF model. 
In this work we explicitly include a decay time scale at the spiking event and we focus on the effects of this time scale on the collective dynamics of a neural network. Instead of modelling potassium and sodium channels, which would add further nonlinearities in the model, we limit ourselves to consider the simplest modification of a LIF model able to keep track of a continuous evolution of the membrane voltage during the spike emission, the \emph{c-LIF}. This model has also the big advantage to be integrable. Interestingly, dynamical regimes similar to those observed in our model have recently been observed  in a non-integrable LIF neural model  with an additional non-linear term that describe the dynamics of the voltage at the spike emission \cite{wang2021}.



\begin{figure}[t]
	\centering
	\includegraphics[width=0.45\textwidth]{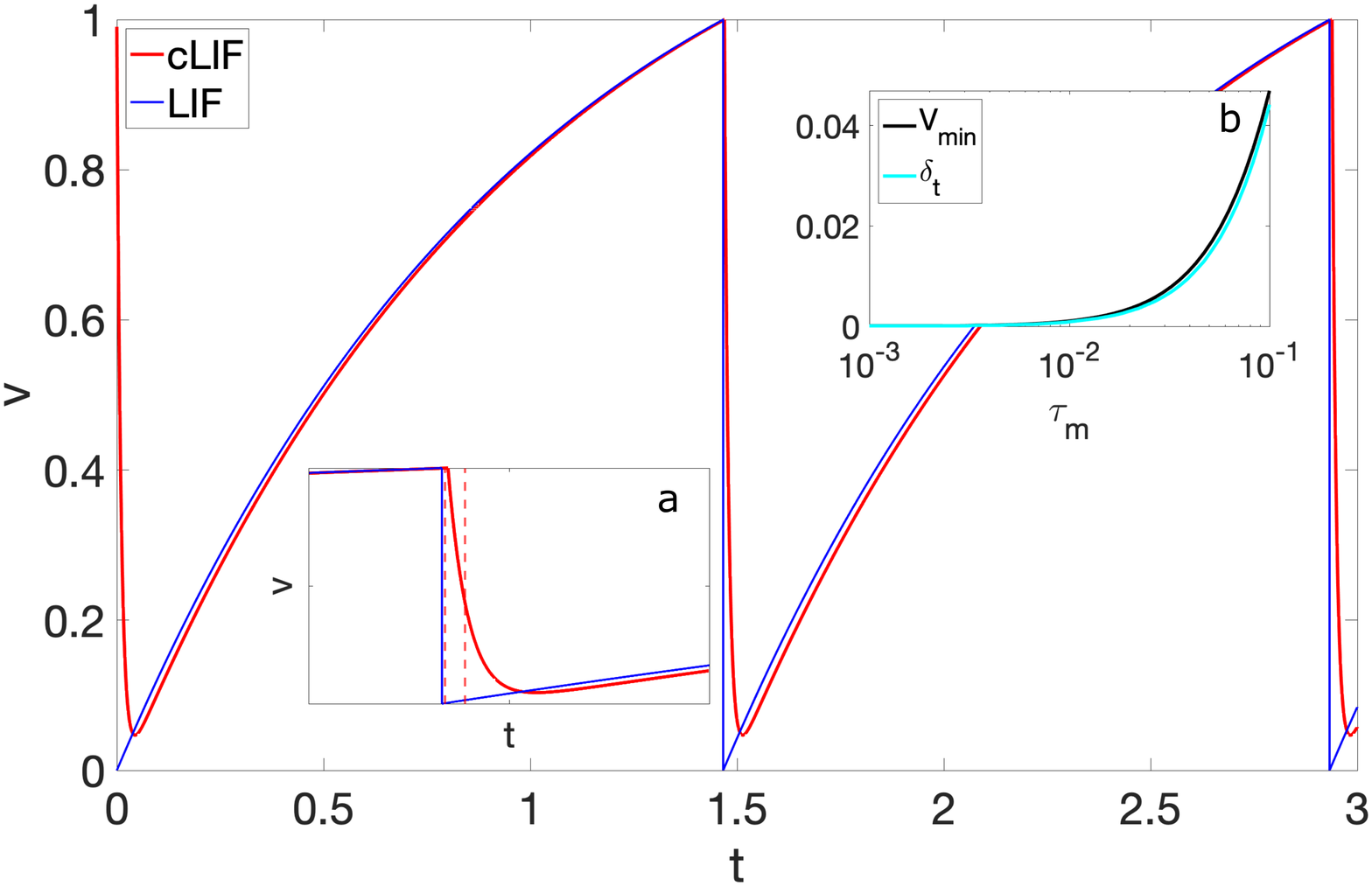}
	\caption{\label{fig:vlib} Time evolution of the rescaled membrane potential $v$ of a free ($g=0$) c-LIF neuron (Eq.~\ref{eq:lifmassadim}) compared to the usual LIF model (Eq. \ref{eq:vk}). Parameters: $a=1.3$, $\tau_{1}=1$ for both models and $\tau_{m}^{2}=10^{-2}$ for the c-LIF one. Inset (a): zoom on the membrane potential decay region of the c-LIF model to be compared with the LIF discontinuity. Vertical dashed lines highlight the decay time $\frac{\tau_{m}^{2}}{\tau_{1}}$ of the isolated c-LIF neuron. Inset (b): in the c-LIF model, plot  of the minimum $V_m$ of the potential $V$ and of the time difference $\delta_t$ between the spiking event and the subsequent minimum of $V$ as a function of $\tau_m$. 	
	}
\end{figure}

The c-LIF model evolves  according to an extension of the LIF equation:
\begin{equation}
	\tau_{m}^{2}\ddot{v}_{i}=-\tau_{1}\dot{v}_{i}+a-v_{i}+gI_i(t).
	\label{eq:lifmassadim}
\end{equation}
We call the term on the left ``inertial term'',  in analogy with  classical mechanics. Eq.~\eqref{eq:lifmassadim} reduces to  LIF Eq.~\eqref{eq:vk} in the limit $\tau_{m} \to 0$. 

With the introduction of the time scale $\tau_{m}$, the equation for the membrane potential is turned into a second order differential equation: as a consequence the state of neuron $i$ is characterized both by the value of its membrane potential $v_{i}$ and by its first derivative $\dot{v}_{i}$. At the spiking times $t_i(n)$, when $v_i(t_i(n))$ reaches the threshold $v_i(t_i(n))=1$, we impose to the membrane potential $v_i(t_i(n))$ to be continuous and we set its first derivative to the negative value:  
\begin{equation}
 \dot{v}_{i}(t_i(n))=-\frac{\tau_{1}}{\tau_{m}^{2}}.
	\label{eq:reset2}
\end{equation}
With this choice, the membrane potential of an \emph{isolated} c-LIF neuron decays at the firing event in a time of order $\frac{\tau_{m}^{2}}{\tau_{1}}$, as shown in Fig. \ref{fig:vlib}. In particular, for $\tau_m\to 0$ we recover the behavior of the LIF model. This is clarified in the inset (b), where we show that for $\tau_m\to 0$ the minimum of the continuous potential tends to zero (i.e. the reset potential of the LIF model) and the time difference $\delta_t$ between the minimum and the spiking event  vanishes.  In analogy with the standard LIF model, we call $ISI_i(n)=t_{i}(n+1)-t_{i}(n)$ the $n$-th inter-spike interval of the c-LIF neuron $i$.

Notice that this rule is not equivalent to a manual reset of the voltage to a value different from zero, known as partial reset rule \cite{bugmann1997role,leng2020common}. Indeed, in our model we are dealing with  interactions that are continuous in time, so that neurons continuously cross the threshold of firing and, as a result, the partial reset rule would have not effects. Therefore the mechanism that we are describing here is actually only due to the decay time and not to a time varying reset rule.

Short-term plasticity tunes the dynamics of neurotransmitter, temporarily affecting neural responses. The widely used TUM model, which we consider here, has been shown to correctly reproduce experimental results for neurotransmitter dynamics in different conditions  \cite{TsodyksUzielMarkram}.  According to the TUM dynamics, the amount of neurotransmitter at each synapse is finite and the state of the $i-$th synapse is determined by three variables: $y_{i}(t)$, $z_{i}(t)$ and $x_{i}(t)=1-y_{i}(t)-z_{i}(t)$,  representing the fraction of neurotransmitters  respectively in the available, active or inactive states. 
The synaptic currents $y_{i}$ evolve in time according to the following dynamical equations:

\begin{eqnarray}
	\dot{y}_{i}(t) & = & -\frac{y_{i}(t)}{\tau_{in}}+u(1-y_{i}(t)-z_{i}(t))S_{i}(t)\label{eq:yk}\\
	\dot{z}_{i}(t) & = & \frac{y_{i}(t)}{\tau_{in}}-\frac{z_{i}(t)}{\tau_{r}},\label{eq:zk}
\end{eqnarray}
where $S_{i}(t)=\sum_{m}\delta(t-t_{i}(m))$ represents the spike train of neuron $i$ ($t_{i}(m)$ is the time when neuron $i$ fires its $m-$th spike). The parameter $u$ regulates the percentage of available resources released at every spike and $\tau_{in}$ and $\tau_{r}$ are the characteristic times of decay of active resources $y_{i}$ and recovery of inactive ones $z_{i}$, respectively. The synapses are depressed (i.e. the effect of the presynaptic neuron on the postsynaptic one is reduced) if the firing rate is high, since neurotransmitters need time to recover from the inactive to the available state.
We assume that all parameters appearing in equations~\eqref{eq:yk} and \eqref{eq:zk} are independent of neuron index and that all efferent synapses of a given neuron follow the same evolution.

We consider an all--to--all network of $N$ c-LIF neurons, where the input current received by neuron i reads:
\begin{equation}   
g I_{i}(t)=g Y(t)=\frac{g}{N}\sum\limits _{j} y_{j},
\label{eq:interaction}
\end{equation}
i.e. every neuron $i$ feels the same mean field $Y$.
Eq.~(\ref{eq:interaction}) can be modified and extended to take into account different topologies, including the inhomogeneities of the system \cite{Burioni2014} and inhibition \cite{bertolotti}.

Eq.s (\ref{eq:lifmassadim}-\ref{eq:interaction}) define our network model which we will study numerically by an event-driven approach \cite{Brette2007,diVoloPRE} (see Appendix for details). To summarize, the model is characterized by four characteristic time scales: $\tau_1$ sets the typical oscillation time of an isolated single neuron, $\tau_m^2/\tau_1$ is the decay time of the action potential, $\tau_{in}$ rules the decay time of synaptic resources and $\tau_r$ their typical recovery time. The constant $a$ determines the shape of the neural oscillation and $g$ is the interaction strength. Setting $\tau_1=1$, our time unit is set to a typical value of the order of $10ms$. Moreover, we expect that $\tau_{in}$ and $\tau_m^2/\tau_1$ are much shorter than $\tau_1$, while $\tau_r$ has a larger value.  In particular, we set the parameters to values that are typically employed in the literature \cite{Pittorino}:  $\tau_{in}=10^{-3}$, $\tau_{r}=10$, $a=1.3$ and $u=0.5$ unless otherwise stated. We vary the values of $\tau_m$ and $g$ to analyse the different dynamical regimes of the network.  We also set $g>0$ since we are only considering excitatory neurons. Notice that we choose these values for the parameters  $a$ and $\tau_{in}$  because they allow to clearly observe and characterise the dynamical effects of the decay timescale over a broad range of values of $\tau_m$ and $g$. Nevertheless, the dynamical regimes that we describe hereafter can be observed even for different values of the parameters. In particular, the avalanche dynamics is observed also at biologically realistic physical values of decay time $\tau_m^2/\tau_1$ (i.e. comparable with the duration of the action potential, see Section 6).


\section{\label{sec:taum_0} The dynamics in the limit $\mathbf{\tau_m \to 0}$}

\begin{figure}[t]
\centering
\includegraphics[width=0.45\textwidth]{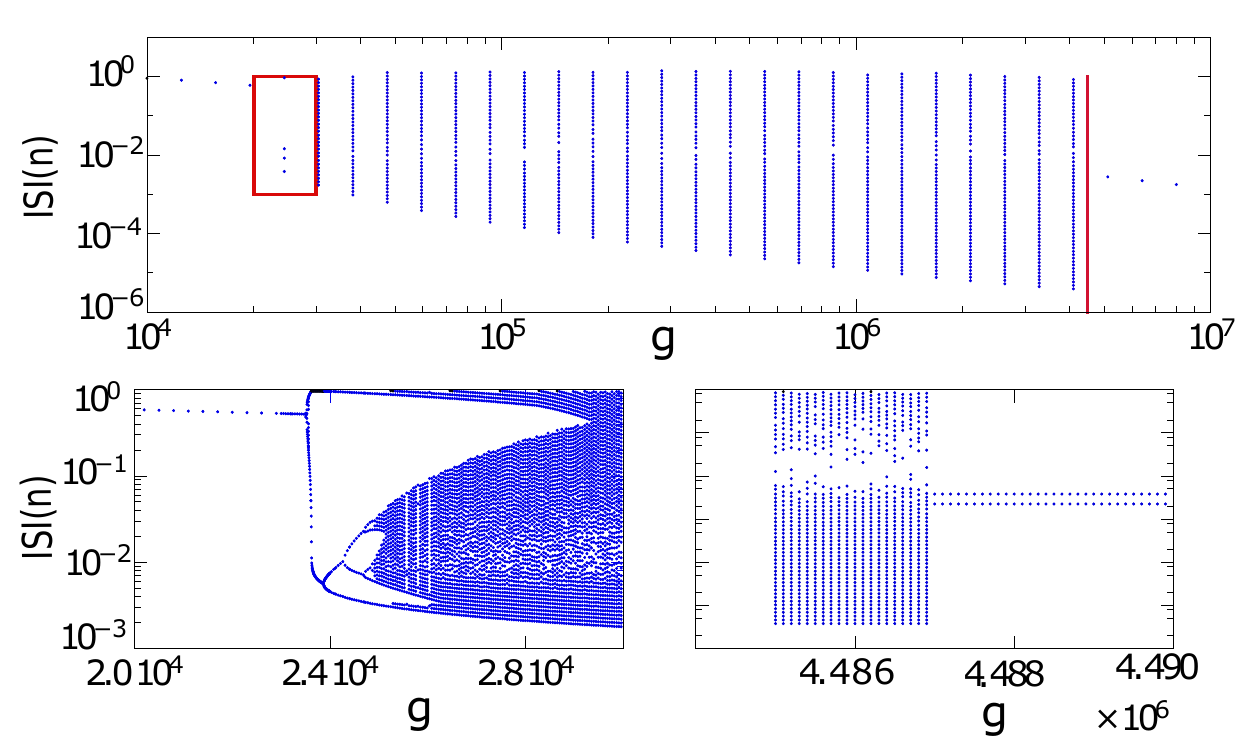}
\caption{\label{fig:pdMF} Bifurcation diagram for the all--to--all LIF model with TUM short-term synaptic plasticity and $\tau_{in} = 10^{-3}$, $\tau_m = 0$.  The attractor for the interspike interval of the network (ISI) is shown in function of the  bifurcation parameter (the coupling constant $g$).
Upper panel: full bifurcation diagram. 
Lower-left panel: zoom on the first transition, revealing a period-doubling transition to chaos.
Lower-right panel: zoom on the second transition. 
Red boxes are the zooming regions.}
\end{figure}

In the limit $\tau_m \to 0$, the c-LIF reduces to the classic LIF model with synaptic plasticity, that we recently investigated in  \cite{Pittorino}. In this limit the dynamics of the all-to-all network  is perfectly synchronous \cite{diVoloPRE} and can be reduced to an effective single neuron equation. Despite its simplicity, the mean field LIF with TUM synaptic plasticity features  an interesting variety of dynamical regimes, depending on the parameters. In particular in Fig.~\ref{fig:pdMF} we show the bifurcation diagram of the interspike intervals for different values of the coupling $g$. 
The bifurcation diagram shows a chaotic regime for intermediate values of $g$.  This regime emerges because of the competition between the slow time scale $\tau_1$, which describes the intrinsic dynamics of the neuron, and the fast interaction time scale $\tau_{in}$ of the synaptic resources. 
In particular,  an heuristic estimate of the effective time scales and interaction strength derived in \cite{Pittorino}  shows that for $\tau_{in}\ll \tau_1$ the dynamics should display a chaotic behavior if  $ K_1 \tau_r/\tau_{in} < g < K_2 (\tau_r\tau_1)/\tau_{in}^{2}$ where the numerical prefactors $K_1$ and $K_2$ are expected to barely depend on the system parameters. In this intermediate regime, indeed, in Eq.~\ref{eq:vk}
the fast interaction term $gI_i(t)=gY(t)$ and the slow evolving terms $a-v(t)$ have the same order of magnitude and one cannot treat one of them perturbatively. As a consequence, the simultaneous  presence of two different timescales gives rise to chaos via a period doubling mechanism.

When introducing disorder (e.g. in the coupling $g$ or, equivalently in the topology of the network), the synchronous chaotic regime is desynchronized and turned into a richer regime with power-law distributed avalanches \citep{Pittorino}.  Neurons with heterogenous interactions have different intrinsic speed and this unpacks the synchronous events observed in the all-to-all model. As a result, the underlying chaotic dynamics in the presence of heterogeneity leads to the emergence of population avalanches.
In the rest of the paper we fix $\tau_{in}=10^{-3}\ll \tau_1=1$, in such a way to span the synchronous, chaotic and periodic regimes by varying $g$. We will show that the introduction of a finite $\tau_m>0$ gives rise to a desynchronizing effect that produces power-law distributed avalanches even in the homogeneous all-to-all network.

\section{\label{sec:dynamics} The role of the time scale $\mathbf{\tau_m}$ in collective dynamics}

The timescale $\tau_m$ has strong effects on the collective dynamics of the homogeneous all-to-all case, which we now investigate. 
We focus on the conservation of the order of neurons spiking events  \cite{kirst2009sequential, timmesiam}, showing that this observable is tightly related to the emergence of bursty dynamics, characterized by avalanches of activity with heavy tailed distribution.  

Let us first note that the mean field LIF model, beside the already mentioned discontinuity of the membrane potential, displays another unphysical feature, that is the conservation of fire ordering. Indeed, at $\tau_m=0$ the firing order is trivially maintained by the dynamical evolution. If neuron $j+1$ fires after neuron $j$, we can define $\Delta v_j(t')=v_j(t')-v_{j+1}(t')$, which is positive at the firing time $t'$ of neuron $j+1$. From the mean field  equations with $\tau_m=0$, we obtain for the dynamics between two firing events of neurons $j+1$ and $j$,  that $\tau_1 \dot{\Delta v_j}(t)=-\Delta v_j(t)$. Hence, $v_j(t)$ remains larger than $v_{j+1}(t)$ until the next firing event where the neuron $j$ is again going to fire before neuron $j+1$, preserving therefore the firing order during the evolution.
On the other hand, for $\tau_m>0$ we have $\tau^2_m \ddot{\Delta v_j}(t)=-\tau_1 \dot{\Delta v_j}(t)-\Delta v_j(t)$, therefore the sign of $\Delta v_j(t)$ is not preserved between firing events and the firing ordering can be violated. 

Hereafter, hence, in the stationary regime we will order neurons according to their firing events and we will study if their firing order is preserved in time. Raster plots are plotted according to this labelling of neurons.
In what follows, we study (otherwise stated) an all--to--all network of $N=500$ neurons, discarding the first $5\cdot 10^6$ firing events in order to reach the stationary regime.

\begin{figure}[t]
	\centering
	\includegraphics[width=0.45\textwidth]{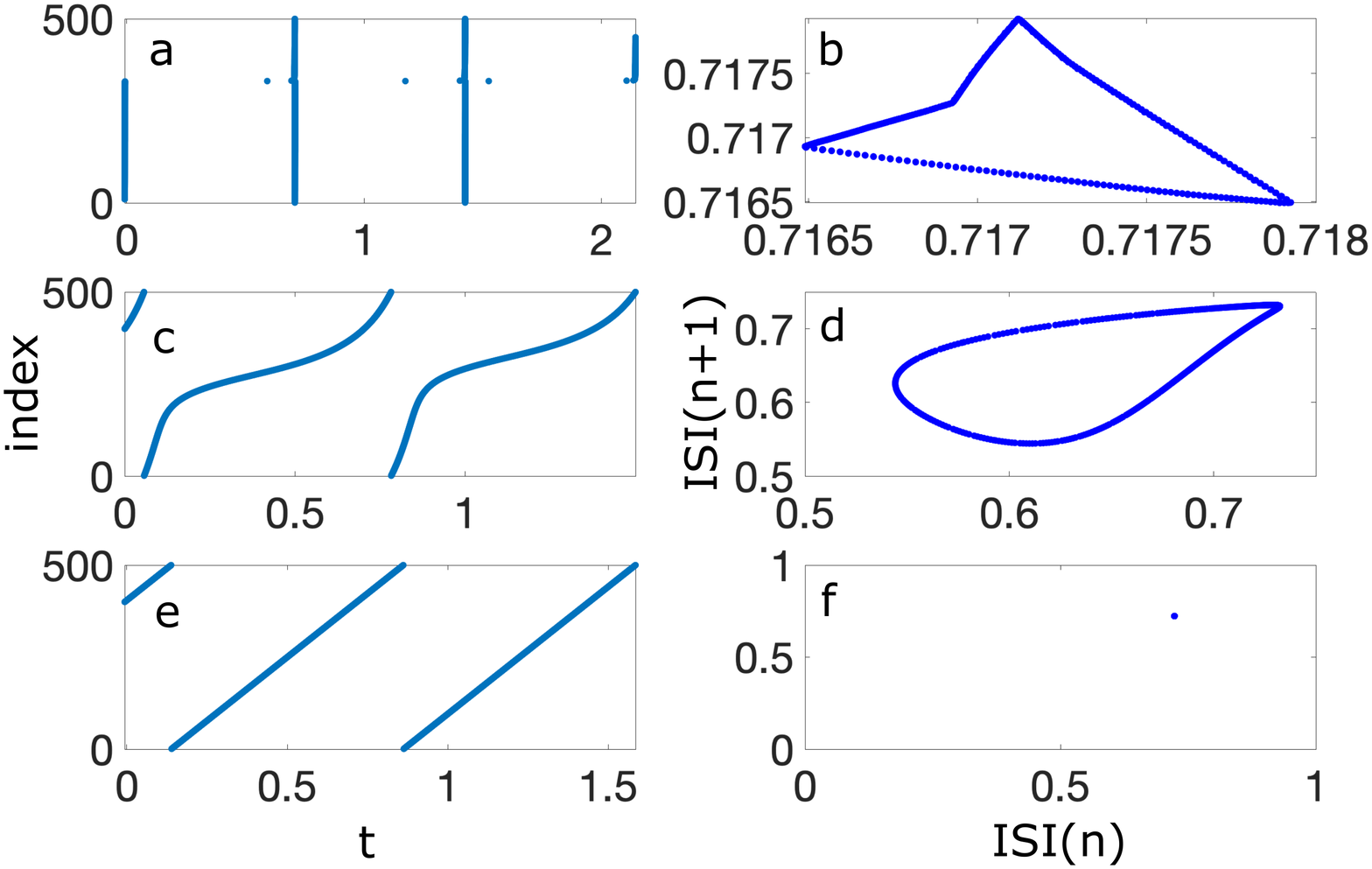}
	\caption{\label{fig:gsmall} Dynamical phases of an all--to--all network of $N=500$ c-LIF neurons with TUM synaptic plasticity at a small value of the coupling, $g=1.5\cdot 10^4$. Left column: raster plots (we have discarded a transient of $t=7\cdot 10^3$). Right column: plot of the $(n+1)-$th interspike interval $ISI(n+1)$ as a function of the $n-$th interspike interval $ISI(n)$ of a single neuron inside the network. Each row in the plot shows the dynamics for a different value of $\tau_m^2$: $\tau_m^2=6.9\cdot 10^{-4}$ for panels a) and b) (quasi-periodic phase); $\tau_m^2=6.8\cdot 10^{-2}$ for panels c) and d) (quasi-periodic phase); $\tau_m^2=1.3\cdot 10^{-1}$ for panels e) and f) (splay state).	
}
\end{figure}

\paragraph{Small coupling}
We first study the case of small coupling $g$, meaning $g<K_1 \tau_r/\tau_{in} $ where the all--to--all network with  $\tau_m=0$ displays a fully synchronous non-chaotic collective dynamics. Fig.~\ref{fig:gsmall} shows that population spiking events are now not completely synchronous as soon as $\tau_m>0$. In particular, in the left column we show the raster plot (for each firing event we plot a dot at the index of the corresponding firing neuron), and in the right column we plot the return map $ISI_{i}(n+1)$ vs. $ISI_i(n)$ for the neuron $i$. Let us remark that even if the interspike interval explicitly depends on the neuron, the curve $ISI_{i}(n+1)$ vs. $ISI_i(n)$ does not  because the system is homogeneous. Hereafter, therefore, we will drop the index $i$.
For very small $\tau_m$ (upper panels) the spikes are almost synchronous, but the $ISI_{n+1}$ vs. $ISI_n$ plot shows the presence of a very small closed attractor, accounting for the presence of a quasi-periodic behavior with small oscillations of the $ISI$ \cite{vree96,serena14}.  For intermediate values of $\tau_m$, middle panels, the quasi-periodic behavior becomes more pronounced and it can be directly observed in the raster plot where firing now is clearly asynchronous. The effect of increasing $\tau_m$ is to lead to a stronger and stronger desynchronization, up to higher values of $\tau_m$ where the dynamics falls in the well known splay state (lower panels), an extremely regular asynchronous regime where neurons firing times are equally spaced \cite{vree96}. 
Remarkably, the firing order is conserved in all these regimes, independently of the value 
of $\tau_m$, as it is shown by the raster plots in Fig.~\ref{fig:gsmall} (we also explicitly check that all pairs of consecutive firing neurons do not exchange their mutual firing order along the dynamics). 

\begin{figure}[t]
	\centering
	\includegraphics[width=0.45\textwidth]{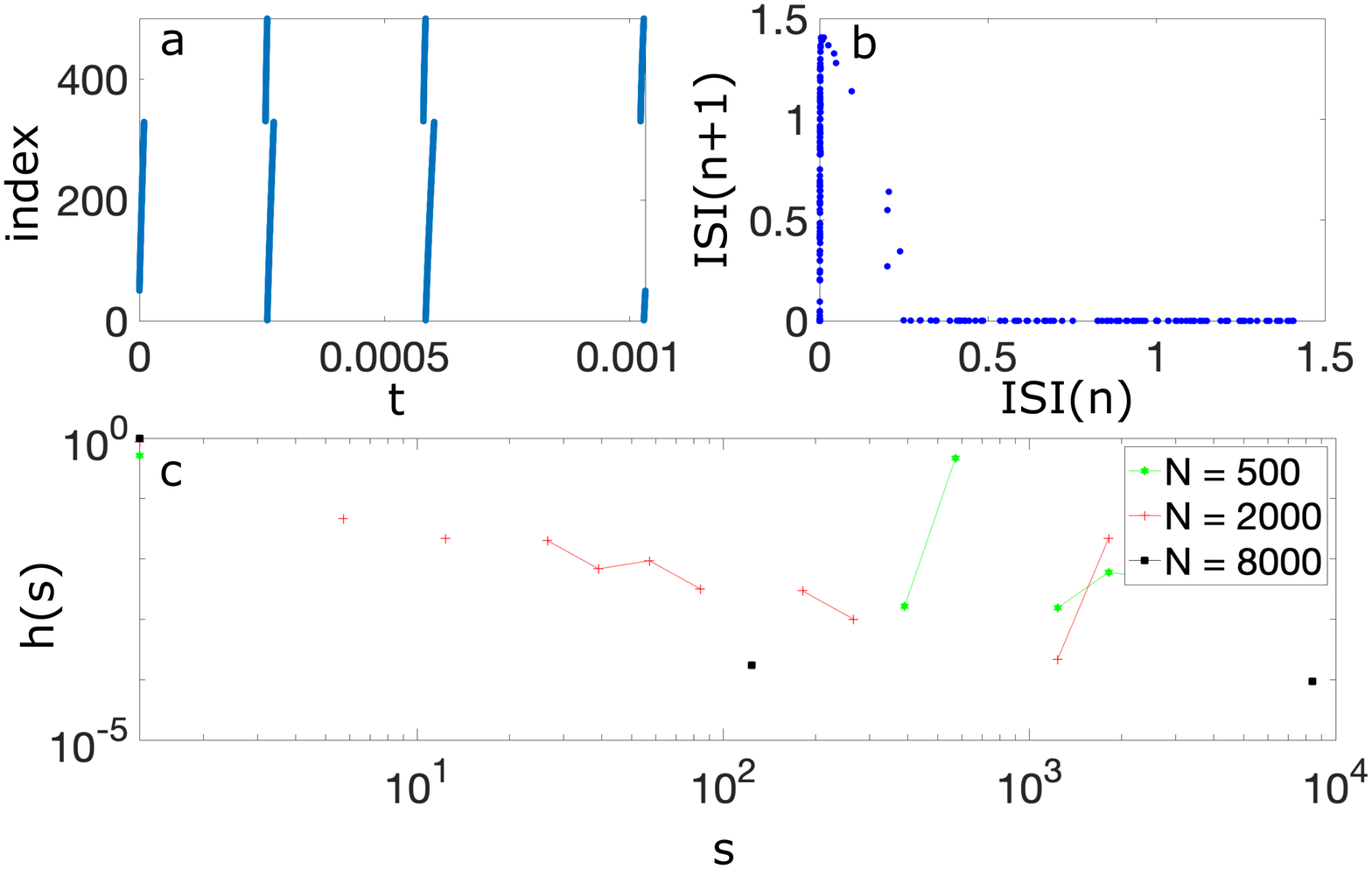}
	\caption{\label{fig:first_taum} Dynamical phase and distribution of events of an all--to--all network of $N=500$ c-LIF neurons with TUM synaptic plasticity at an intermediate value of the coupling, $g=1.0\cdot 10^5$ and for a low value of $\tau_m^2 =1.4\cdot 10^{-5}$. Upper left, panel a): raster plot (we have discarded a transient of $t=2\cdot 10^3$). Upper right, panel b): plot of the $(n+1)-$th interspike interval $ISI(n+1)$ as a function of the $n-$th interspike interval $ISI(n)$ of a single neuron in the network. As shown by this latter panel the dynamics is chaotic, but there is no firing order symmetry breaking.
Lower panel c): in correspondence to this dynamical phase characterized by the preservation of neurons firing order, the distribution of events is not broad (at increasing values of $N = \{500, 2000, 8000\}$).
	 }
\end{figure}

\begin{figure}[t]
	\centering
	\includegraphics[width=0.45\textwidth]{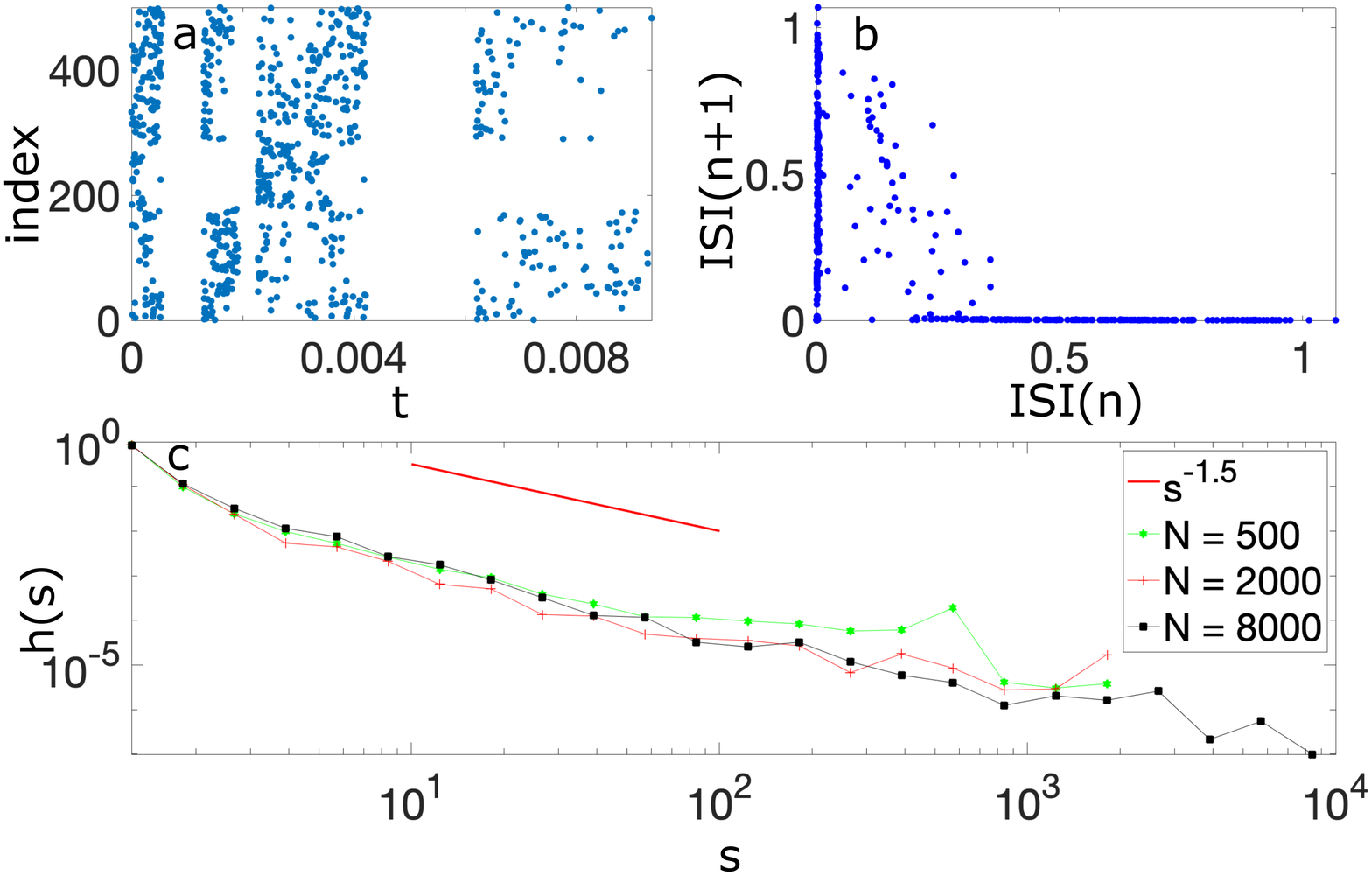}
	\caption{\label{fig:second_taum} Dynamical phase and distribution of events of an all--to--all network of $N=500$ c-LIF neurons  with TUM synaptic plasticity at an intermediate value of the coupling, $g=1.0\cdot 10^5$ and for an intermediate value of $\tau_m^2 =7.0\cdot 10^{-4}$. Upper left, panel a): raster plot (we have discarded a transient of $t=2\cdot 10^3$). Upper right, panel b): plot of the $(n+1)-$th interspike interval $ISI(n+1)$ as a function of the $n-$th interspike interval $ISI(n)$ of a single neuron inside the network. As shown by this latter panel the dynamics is chaotic, and in this phase there is firing order symmetry breaking.
Lower panel c): in correspondance to this dynamical phase characterized by the breaking of neurons firing order, there is a broad distribution of events (at increasing values of $N = \{500, 2000, 8000\}$).
	}
\end{figure}

\begin{figure}[H]
	\centering
	\includegraphics[width=0.45\textwidth]{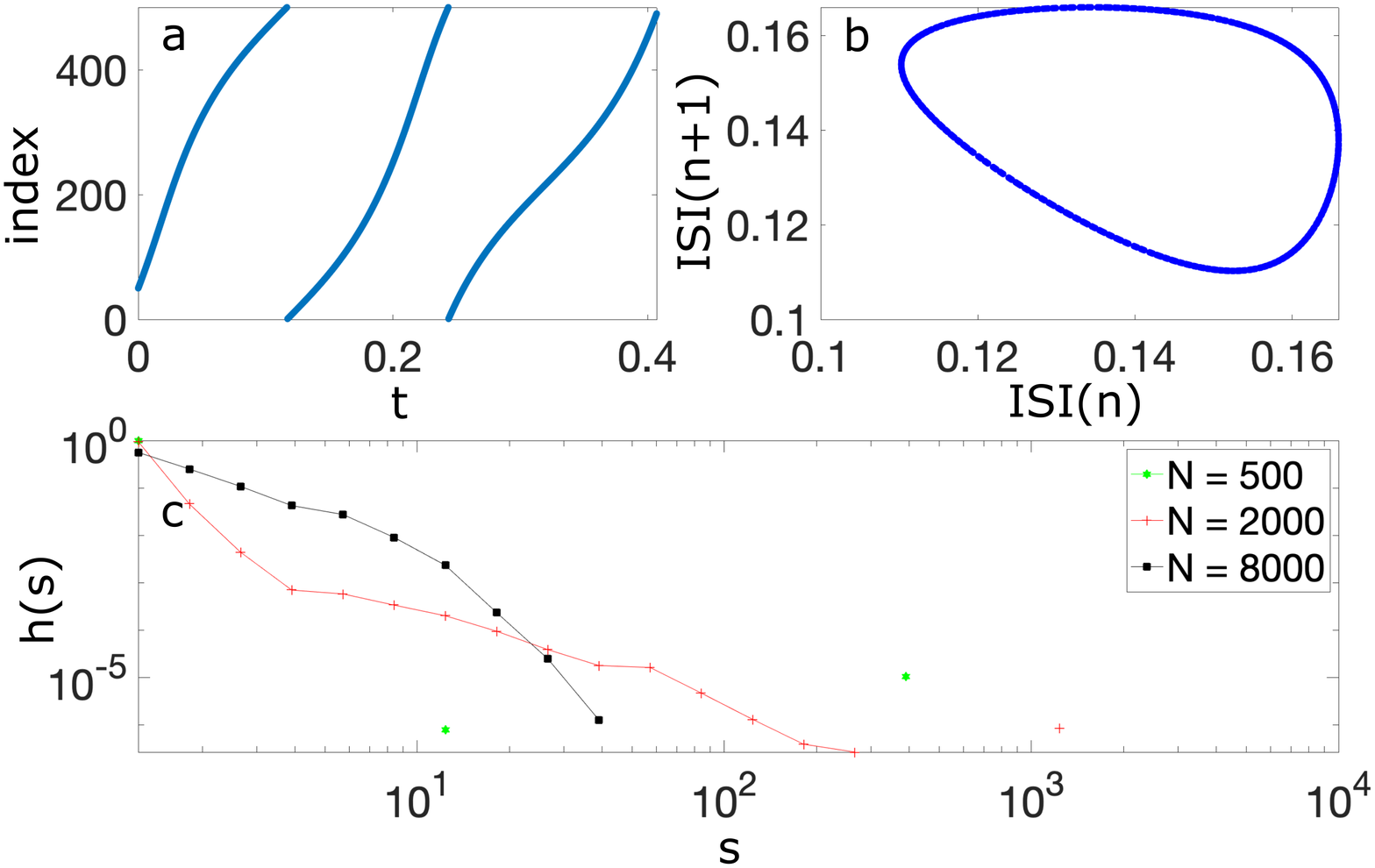}
	\caption{\label{fig:third_taum} Dynamical phase and distribution of events of an all--to--all network of $N=500$ c-LIF neurons with TUM synaptic plasticity at an intermediate value of the coupling, $g=1.0\cdot 10^5$ and for $\tau_m^2 =9.5\cdot 10^{-3}$. Upper left, panel a): raster plot (we have discarded a transient of $t=2\cdot 10^3$). Upper right, panel b): plot of the $(n+1)-$th interspike interval $ISI(n+1)$ as a function of the $n-$th interspike interval $ISI(n)$ of a single neuron inside the network. As shown by this latter panel the dynamics is quasi-periodic, and the firing order becomes preserved again.
Lower panel c): in correspondance to this dynamical phase characterized by the preservation of neurons firing order, the distribution of events is not broad (at increasing values of $N = \{500, 2000, 8000\}$).
	}
\end{figure}

\paragraph{Intermediate coupling}
The  $\tau_m=0$ case for intermediate values of $g$ is characterized by  synchronous chaotic dynamics, and also in this regime  a time constant $\tau_m>0$ induces a desynchronization. At very small $\tau_m$  (see Fig.~\ref{fig:first_taum} upper panels), the synchronous burst structure is only slightly modified. One can indeed observe  the presence of clearly separated almost synchronous bursts where each neuron fires exactly once. The firing order in this regime is preserved during the evolution. By increasing $\tau_m$ (see  Fig.~\ref{fig:second_taum} upper panels)  bursts are so broadened that they overlap. The ${ISI}(n+1)$ vs. ${ ISI}(n)$ plot shows the presence of a non-trivial attractor, pointing at an underlying chaotic dynamics, where the firing order is not preserved. At larger $\tau_m$ the system enters into a quasi-periodic regime, clearly visible from the one-dimensional closed attractor in the ${ ISI}(n+1)$ vs. ${ISI}(n)$ in Fig.~\ref{fig:third_taum}). Here, the neurons firing order is again preserved. Finally, at even larger $\tau_m$ the dynamics becomes asynchronous and the system falls in the splay state (see Fig.~\ref{fig:fourth_taum}).

\paragraph{Large coupling}
In the case of large coupling, i.e. for $g>K_2 (\tau_r\tau_1)/\tau_{in}^{2}$ the dynamics at $\tau_m=0$ is again synchronous and periodic. However now a positive value of $\tau_m>0$ fully destabilizes the synchronous dynamics and we enter in an asynchronous regime. Indeed, by perturbing the synchronous state at $\tau_m=0$ the dynamics reaches the asynchronous splay state for any value of  $\tau_m>0$; in particular, the smaller $\tau_m^2$, the longer it takes to the system to desynchronize.
Clearly, in the  splay state, the firing order is conserved for any values of $\tau_m$.

\section{\label{sec:symmetry} Order symmetry breaking and broad distributions of events}

As already observed, in the regime described in Fig.~\ref{fig:second_taum} fire ordering is not conserved by the dynamical evolution. Two conditions seem to be necessary to violate the ordering: first the fully synchronized dynamics at $\tau_m=0$ should be chaotic, second a desynchronizing mechanism is needed allowing for the overlap of different bursts, and this can be caused by disorder in the couplings \cite{Pittorino} or by the inertial time scale $\tau_m$. In the latter case, the reset time of a neuron $\tau_m^2/\tau_1$ provides an estimate of the duration of a broadened burst and, as far as this time scale is much smaller than the minimal time between two consecutive bursts, no overtaking is possible. In this perspective, we verify that the minimal ${ ISI}_{ min}$ measured at $\tau_m=0$ (i.e. the minimal $ISI$ values appearing in the {\em vertical bars} in the bifurcation diagram in Fig.~\ref{fig:pdMF}) provides the order of magnitude of $\tau_m^2/\tau_1=\tau^*$ at which we start to observe a violation of firing order in the presence of $\tau_m$. For different values of $g$ in the chaotic regime, these values coincide with what is observed in Fig.~\ref{fig:pdMF}. Measured values of ${ ISI}_{ min}$ and $\tau^*$ as a function of $g$ are reported in Table~\ref{table:1}.
\begin{figure}[t]
	\centering
	\includegraphics[width=0.45\textwidth]{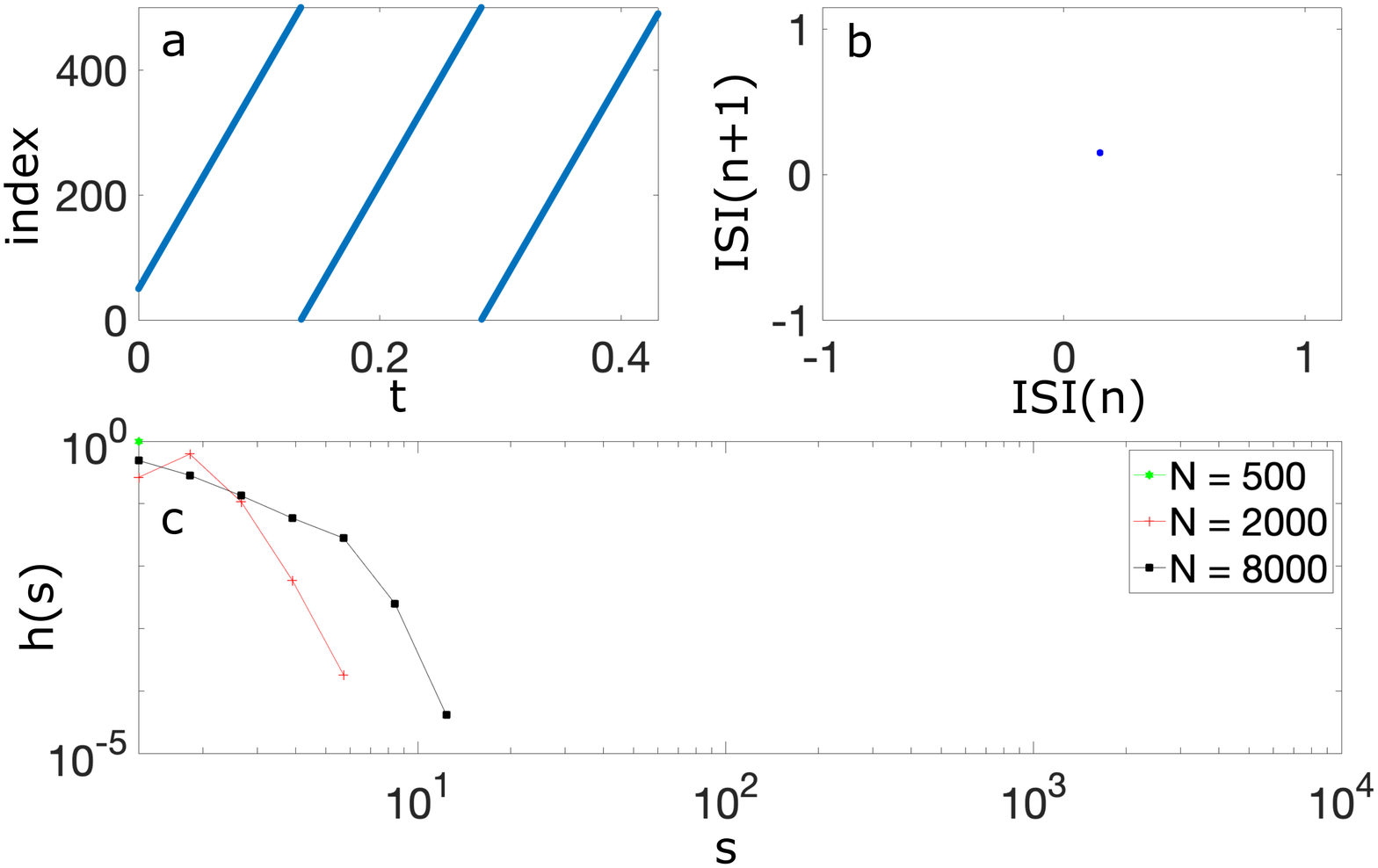}
	\caption{\label{fig:fourth_taum}  Dynamical phase and distribution of events of an all--to--all network of $N=500$ c-LIF neurons with TUM synaptic plasticity at an intermediate value of the coupling, $g=1.0\cdot 10^5$ and for a high value of $\tau_m^2 =1.8\cdot 10^{-2}$. Upper left, panel a): raster plot (we have discarded a transient of $t=2\cdot 10^3$). Upper right, panel b): plot of the $(n+1)-$th interspike interval $ISI(n+1)$ as a function of the $n-$th interspike interval $ISI(n)$ of a single neuron inside the network. As shown by this two panels the dynamics is a splay state, in which the firing order is preserved.
Lower panel c): in the splay state, the distribution of events is not broad (at increasing values of $N = \{500, 2000, 8000\}$).
	}
\end{figure}

\begin{table}
    \centering
    \begin{tabular}{|c|c|c|}
        \hline        
        $g$ & $\tau^*$ & ${ ISI}_{ min}$ \\
        \hline \hline
        $10^5$ & $9.7 \cdot 10^{-5}$ & $1.5\cdot 10^{-4}$ \\
        \hline
        $4.6\cdot 10^5$ & $2.6\cdot 10^{-5}$ & $3.1\cdot 10^{-5}$\\
        \hline
        $10^6$ & $1.4\cdot 10^{-5}$ & $1.0\cdot 10^{-5}$\\
        \hline
    \end{tabular}
    \caption{Measured values of the minimum $ISI$ value ${ ISI}_{ min}$ at $\tau_m=0$ compared with the value $\tau^*=\tau_m^2/\tau_1$ at which firing order begins to be violated in the model with $\tau_m>0$, for three different values of the coupling $g$.} 
    \label{table:1}
\end{table}

\begin{figure}[t]
	\centering
	\includegraphics[width=0.45\textwidth]{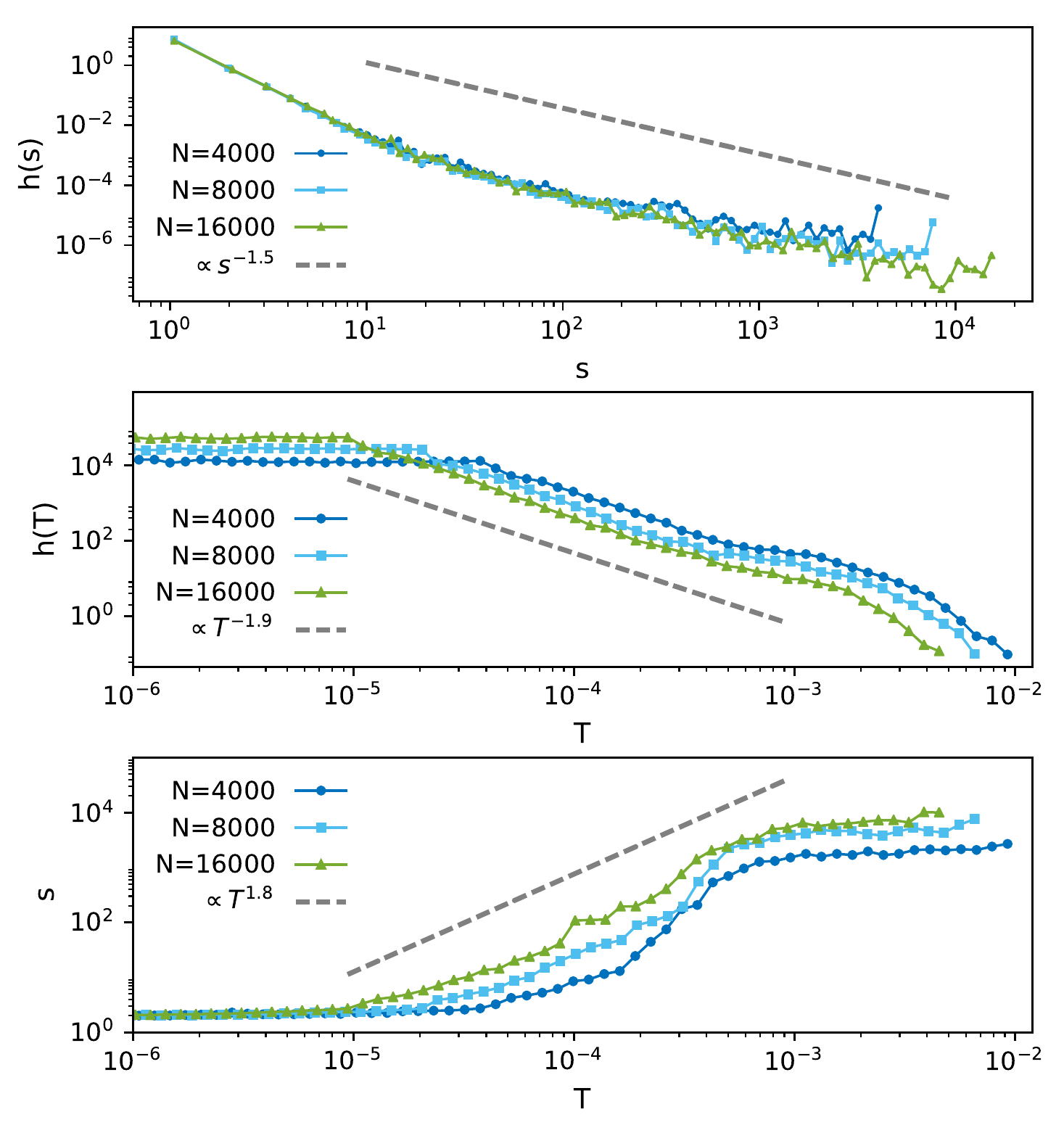}
	\caption{\label{fig:scaling} (color online) Size (s) and duration (T) distributions of all--to--all networks of cLIF neurons endowed with TUM synaptic plasticity for $g=1.0\cdot 10^5$ and $\tau_m^2 =7.0\cdot 10^{-4}$. Upper panel: the size distribution is a power-law with exponent $\tau_s=1.5$. Middle panel: duration distribution, showing a power-law with $\tau_T=1.9$. Lower panel: average size for fixed duration, showing a power law with exponent $\tau=1.8$. In all cases the power-law scaling holds for increasing values of $N = \{4000, 8000, 16000\}$). The values of the exponents are fitted over the range of the corresponding dashed line and satisfy the relation $\tau=\frac{\tau_T-1}{\tau_s-1}$, as expected for a system near criticality.}
\end{figure}
In the dynamical regime where fire ordering is broken, we now measure the size and duration distribution of avalanches. We apply the avalanche definition adopted in experimental settings \cite{Beggs03122003}.  An \emph{event} or avalanche is defined as a set of consecutive neuronal spikes such that all interspike intervals $t(spike_{n+1})-t(spike_{n})$ are smaller than a threshold $\delta$ \cite{Beggs03122003}, setting $\delta$ of the same order of magnitude of the average interspike interval of the network. The size $s$ of an event is then defined as the number of spikes it contains. Analogously the duration of an event is the time elapsed between the first and the last spikes of the event.  We then consider the distributions of sizes $h(s)$, durations $h(T)$ and mean avalanche size $s$ as a function of the duration $T$, as shown in Fig.~\ref{fig:scaling}.

Figures~\ref{fig:first_taum}-\ref{fig:scaling} summarize our results.  A broad distribution in the event size is observed only in the regime where the firing order is not conserved. For small values of $\tau_m^2$ (Fig.~\ref{fig:first_taum}) indeed we observe an irregular dynamics but the neurons fire in quasi-synchronous bursts and typically one observes only large events while small events are rare. In the regular quasi-periodic regime (Fig.~\ref{fig:third_taum}) and in the splay state (Fig.~\ref{fig:fourth_taum}) no large events are present and a cut-off in the  characteristic size $h(s)$ is observed. Instead, the overlap of the different bursts and the overtaking in firing order that characterizes the regime in Fig~\ref{fig:second_taum} gives rise to a broad distribution in $h(s)$. 
In particular, such distribution is compatible with a power law $h(s)\sim s^{-\tau_s}$
with $\tau_s \simeq 1.5$. The  distribution does not appear to depend significantly on the threshold $\delta$ and it scales with $N$ to higher values of $s$ as $N$ is increased, since a larger $N$ allows the presence of
larger avalanches. Analogously, as shown in Fig.~\ref{fig:scaling}, the distribution of avalanches durations $T$ shows the same power law behavior, with exponent $\tau_T=1.9$. Interestingly, the exponents satisfy the scaling relation $\tau=\frac{\tau_T-1}{\tau_s-1}$, where $\tau=1.8$ is the exponent for the average size as a function of fixed duration, as expected for a system near criticality \citep{munoz99}.

\section{\label{sec:synchronization}Synchronization and the Kuramoto parameter}
In this section we consider the Kuramoto parameter \cite{kura_original} as a measure of the synchronicity level of the neurons, to characterize the different dynamical regimes of the model. The Kuramoto parameter reads:
\[
	R(t)=\frac{1}{N}\left|\sum\limits _{i=1}^{N}e^{i\phi_{i}(t)}\right|
\]
where $\phi_{i}(t)$ is the phase of neuron $i$ at time $t$:
\[
	\phi_{i}(t)=2\pi\frac{t-t_{i}(m)}{t_{i}(m+1)-t_{i}(m)}
\]
and $t_{i}(m)$ is the $m-$th spike of neuron $i$ and $t\in[t_{i}(m),\:t_{i}(m+1)]$.
The Kuramoto parameter $R$ and its average in time $\langle R\rangle$ take values
in the interval $[0,1]$, ranging from an asynchronous system at $\langle R\rangle=0$ to a perfectly synchronous one at $\langle R\rangle=1$.

\begin{figure}[H]
	\centering
	\includegraphics[width=0.45\textwidth]{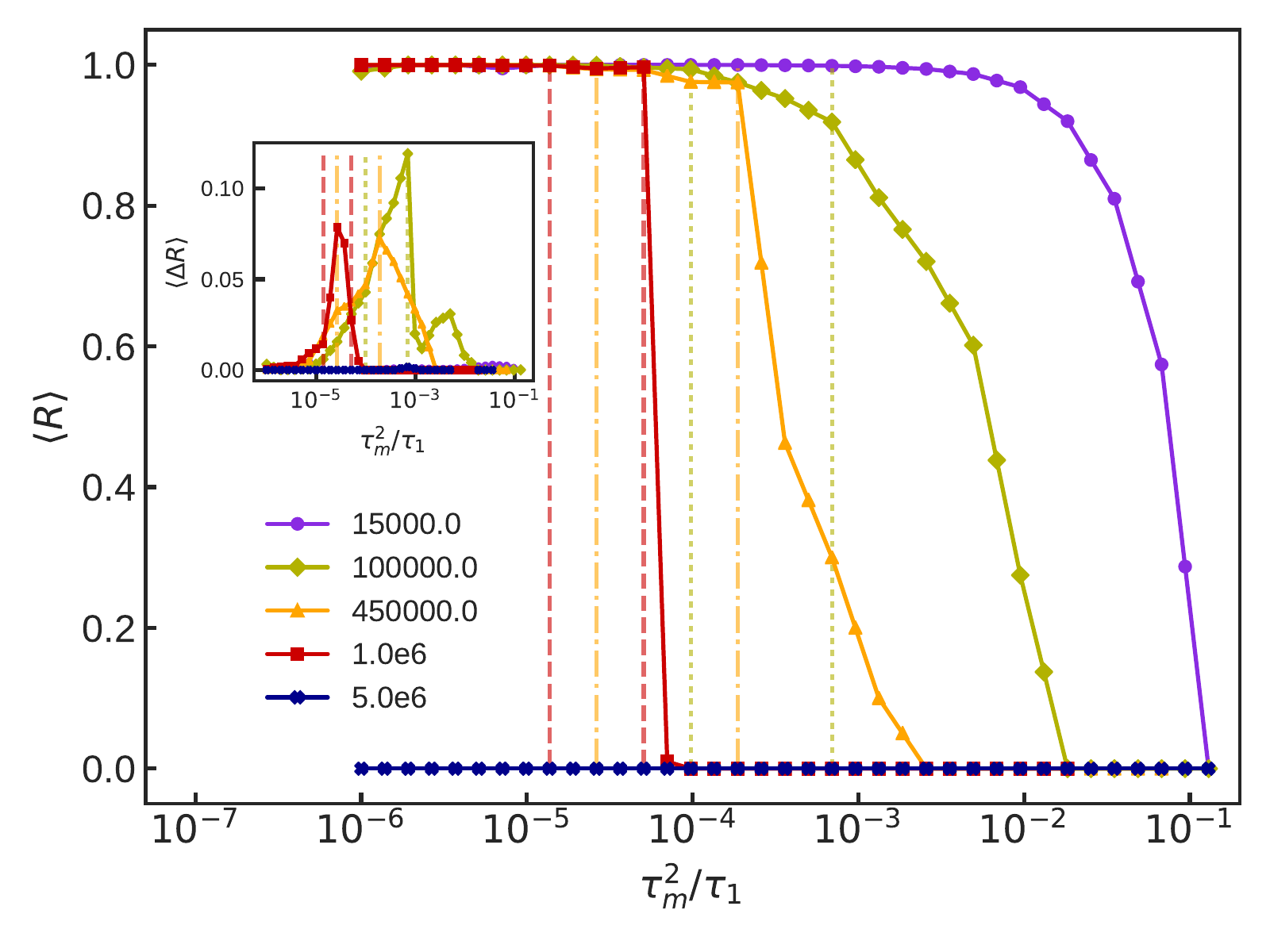}
	\includegraphics[width=0.45\textwidth]{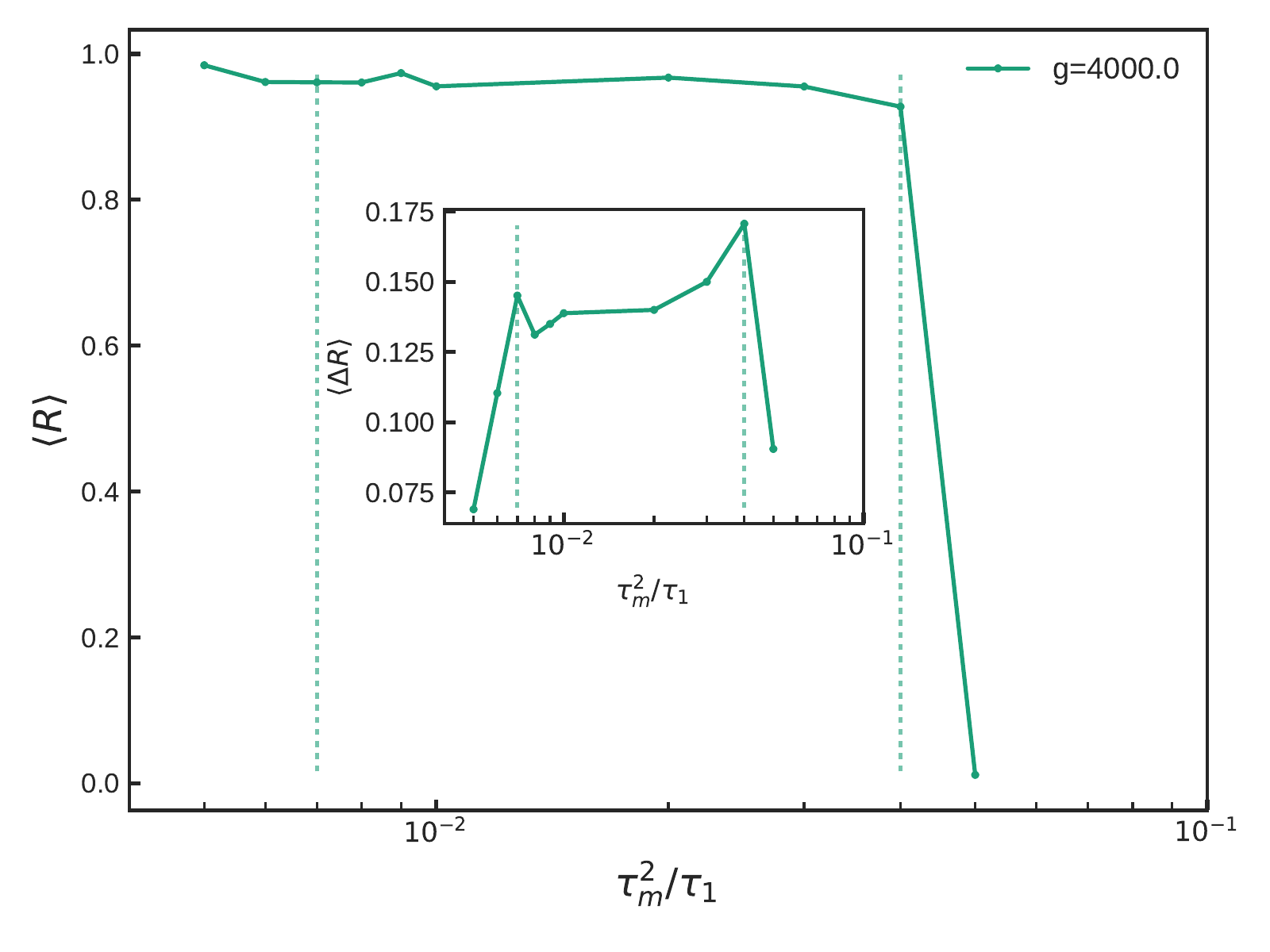}
	\caption{\label{fig:desinc} (color online) Upper panel: the time-averaged value $\langle R \rangle$ of the Kuramoto parameter and of its standard deviation $\langle \Delta R \rangle$ as a function of $\frac{\tau_{m}^{2}}{\tau_1}$ for different values of the coupling parameter $g$ spanning the whole quasi-synchronous to asynchronous transition (each curve corresponds to a value of $g$ according to the legend). Here $a=1.3$ and $\tau_{in}=10^{-3}$. Regions of the parameter space characterized by exchanges in neuron firing order and broad distribution of events are delimited by vertical lines (corresponding to the values of $g$ according to their color). Lower panel: Kuramoto parameter and its standard deviation (inset) for $a=1.01$, $\tau_{in}=10^{-2}$, $g=4\cdot 10^{2}$. Notice that for these value of the parameters, we observe the bursty phase (delimited by the dashed vertical lines) up to a value of $\tau_m^2=4\cdot 10^{-2}$, which corresponds to a decay time of $0.4$ms in physical units (if we consider $\tau_1=10$ms).
	}
\end{figure}

Figure~\ref{fig:desinc} shows a summary of the different behaviors. Continuous lines corresponds to  the Kuramoto parameter and dashed vertical lines delimit the region where overtaking in neuron dynamics are observed 
and broad distributions of avalanches are present. At small $g=1.5\cdot 10^4$ no overtaking is present. In the quasi-periodic regime the average Kuramoto parameter $\langle R \rangle$ is positive, then a transition to the splay state asyncronous regime with $\langle R \rangle=0$ is observed  at larger $\tau_m$. For the intermediate values of $g$ at large $\tau_m$ we also observe a transition from the quasi periodic regime to the splay state where $\langle R \rangle=0$. At lower $\tau_m$ vertical dashed lines delimit the region where exchanges in neuron firing order occur and avalanche sizes and durations are in the scaling regime. Interestingly in this regime the average Kuramoto parameter is quite large $.918 \lessapprox \langle R \rangle \lessapprox .995$ signaling the presence of almost synchronous events even if the sizes of such events fluctuate. Then, for even smaller $\tau_m$, the dynamics is chaotic and almost synchronous  as indicated by  the Kuramoto parameter ($\langle R \rangle \approx 0.999$). Finally for larger $g=5\cdot 10^6$ the average Kuramoto parameter $\langle R \rangle$ vanishes for any value of $\tau_m>0$: this means that the system is in an asynchronous splay state.

We show the standard deviation $\langle \Delta R \rangle$ of the Kuramoto parameter in the inset of Fig.~\ref{fig:desinc}. We observe as expected $\langle \Delta R \rangle \approx 0$ in the quasi-periodic and splay states, while we measure non-negligible values $.025 \lessapprox \langle \Delta R \rangle \lessapprox .12$, indicating strong fluctuations of the Kuramoto parameter, in correspondence to the region where exchanges in neuron firing and broad distributions of avalanches are present. 
Strong fluctuations may be present also at higher values of $\tau_m^2$, after the firing crossing region, where a quasi-periodic phase with a complex two-dimensional orbit is observed. This phase is characterised by lower synchronization, $\langle R \rangle \lessapprox 0 .8$. 
A similar scenario concerning the fluctuations of this order parameter has been reported in \cite{Pittorino}.

The dynamical regime that we observe and characterize is expected to be present also for more realistic value of the involved timescales. The lower panel of Figure~\ref{fig:desinc} shows that, for different values of the parameters, the bursty dynamical regime with avalanches is observed even at larger decay times $\tau_m^2/\tau_1$, which are comparable with the realistic duration of the action potential. In particular, the bursty phase is present up to values of $\tau_m^2=4\cdot 10^{-2}$, which corresponds to a decay time of $0.4$ms in physical units if the membrane time constant  is set to $\tau_1=10$ms.

\section{\label{sec:conclusions} Conclusions}

We have investigated the synchronization dynamics of a fully connected neural network with short-term synaptic plasticity with a new membrane potential time scale in the equations governing the single neuron dynamics. The c-LIF model that we study here avoids unphysical discontinuities in the membrane potential characterizing the usual LIF neurons, while still being integrable. We have shown that if the time constant $\tau_{m}$ related to the new term is not too large the system maintains its variety of dynamical phases. For small values of the coupling $g$ the network is in a quasi-periodic, quasi-synchronous state, while for large values the system is completely asynchronous. 
Interestingly, the desynchronizing effect of this time scale induces a more complex regime for intermediate values of $g$, in the presence of a chaotic dynamics. The bursting regime is characterized by a broad and robust distribution of event sizes and durations, even in a pure mean field model with homogenous couplings. We have numerically measured the exponents of these distributions and shown that they satisfy a scaling relation. The onset of this regime is related to the order symmetry breaking of the spike times of single neurons, which produces a non trivial dynamical regime even in the absence of disorder or heterogeneity.  
Notice that in the context of phase oscillators the presence of a mass term has been found to give rise to interesting dynamical regimes with breaking of symmetry in an homogeneous system, called chimeras \cite{olmi15}. In our neural model the mass term emerges naturally from the introduction of a continuous dynamics of the membrane voltage and is the essential factor giving rise to chaotic dynamics and avalanches as observed in experimental data of neuronal network.




\appendix
\section{Event-driven map}

It is convenient for numerical simulations to transform the set of equations \eqref{eq:lifmassadim}, \eqref{eq:yk}, \eqref{eq:zk} and \eqref{eq:interaction} into an event-driven map \cite{Brette2007,diVoloPRE}, which does not require the numerical integration of the neural equations. Indeed, it is possible to find the exact analytical solution of the LIF and c-LIF equations as they are linear between two consecutive spikes of the network. The idea is to find the analytical solution to these equations from the time immediately after the last spike of the network, $t_{n}$, to the time immediately before the next spike of the network, $t_{n+1}$. As long as $t_{n}<t<t_{n+1}$ the last term on the right in eq.~\eqref{eq:yk} does not contribute.

In order to make the equations readable, we define the following coefficients:
$v_{i,n}=v_{i}(t_{n})$, $\dot{v}_{i,n}=\dot{v}_{i}(t_{n})$,
$Y_{n}=g\tilde{k}Y(t_{n})$, $\Delta=t-t_{n}$, $\alpha=\sqrt{\tau_{1}^{2}-4\tau_{m}^{2}}$,
$\beta=\frac{\tau_{1}+\alpha}{2\tau_{m}^{2}}$, $\gamma=\frac{\tau_{1}-\alpha}{2\tau_{m}^{2}}$,
$T_{1}=\tau_{m}^{2}+\tau_{in}(\tau_{in}-\tau_{1})$, 
and we obtain the following form for the event-driven map of the c-LIF model:

\begin{widetext}
	\begin{align}
	v_{i}\left(t\right)= & a+\frac{2\tau_{m}^{2}}{\tau_{in}^{2}\alpha^{3}-\left(\tau_{1}\tau_{in}-2\tau_{m}^{2}\right)^{2}\alpha}\bigg\{-e^{-\frac{\Delta}{\tau_{in}}}2Y_{n}\tau_{in}^{2}\alpha\nonumber \\
 	& +e^{-\beta\Delta} \Big[ -aT_{1}\left(\tau_{1}-\alpha\right)+Y_{n}\tau_{in}\left(2\tau_{m}^{2}+\tau_{in}\left(-\tau_{1}+\alpha\right)\right)\nonumber +T_{1}\left(2\tau_{m}^{2}\dot{v}_{i,n}+v_{i,n}\left(\tau_{1}-\alpha\right)\right) \Big]\nonumber \\
 	& +e^{-\gamma\Delta} \Big[ aT_{1}(\tau_{1}+\alpha)+Y_{n}\tau_{in}\left(-2\tau_{m}^{2}+\tau_{in}\left(\tau_{1}+\alpha\right)\right) \nonumber -T_{1}\left(2\tau_{m}^{2}\dot{v}_{i,n}+v_{i,n}\left(\tau_{1}+\alpha\right)\right) \Big] \bigg\}
 	\label{eq:vmap}
	\end{align}
	
	\begin{align}
	\dot{v}_{i}\left(t\right)= & \frac{2\tau_{m}^{2}}{\tau_{in}^{2}\alpha^{3}-\left(\tau_{1}\tau_{in}-2\tau_{m}^{2}\right)^{2}\alpha}\Big\{ e^{-\frac{\Delta}{\tau_{in}}}2Y_{n}\tau_{in}\alpha\nonumber \\
 	& -e^{-\beta\Delta}\left[-2aT_{1}+Y_{n}\tau_{in}\left(\tau_{1}-2\tau_{in}+\alpha\right)+T_{1}\left(2v_{i,n}+\dot{v}_{i,n}\left(\tau_{1}+\alpha\right)\right)\right]\nonumber \\
 	& -e^{-\gamma\Delta}\left[2aT_{1}+Y_{n}\tau_{in}\left(-\tau_{1}+2\tau_{in}+\alpha\right)-T_{1}\left(2v_{i,n}+\dot{v}_{i,n}\left(\tau_{1}-\alpha\right)\right)\right]\Big\}
	\end{align}
\end{widetext}

\begin{equation}
	y_{i}(t)=y_{i}(t_{n})e^{-\frac{\Delta}{\tau_{in}}}
\end{equation}

\begin{equation}
	z_{i}(t)=z_{i}(t_{n})e^{-\frac{\Delta}{\tau_{r}}}+\frac{\tau_{r}}{\tau_{r}-\tau_{in}}y_{i}(t_{n})(e^{-\frac{\Delta}{\tau_{r}}}-e^{-\frac{\Delta}{\tau_{in}}})
\end{equation}

The time of the next spike of the network is $t_{n+1}=\min_{i}\left\{ t_{i}\right\} $ where the set $\left\{ t_{i}\right\} $ is determined by imposing $v_{i}(t_{i})=1$ in eq. \eqref{eq:vmap}. Once that $t_{n+1}$ is known it is possible to update the state variables of the i-th neuron:
$v_{i}(t_{n})\leftarrow v_{i}(t_{n+1})$, $\dot{v}_{i}(t_{n})\leftarrow \dot{v}_{i}(t_{n+1})$,
$y_{i}(t_{n})\leftarrow y_{i}(t_{n+1})$, $z_{i}(t_{n})\leftarrow z_{i}(t_{n+1}).$

Moreover, for the particular set of neurons $r$ spiking at time $t_{n+1}$ the value of $\dot{v}$ has to be reset to $-\frac{\tau_{1}}{\tau_{m}^{2}}$:
\begin{equation}
	\dot{v}_{r}=-\frac{\tau_{1}}{\tau_{m}^{2}} \label{eq:reset}
\end{equation}
and the activated resources $ux$ need to be added to $y$ and consequently to the global field $Y$. See also \cite{diVoloPRE, Pittorino} for details on the event-driven simulation of the LIF model. By repeating these steps we obtain the full evolution of the system.


\bibliographystyle{apsrev4-1}
\bibliography{biblioneuro}

\end{document}